\documentstyle [12pt, psfig]{article} 
\title{Light-induced conversion of nuclear spin isomers of molecules}
\author{A.M.Shalagin\thanks{
E-mail: shalagin@iae.nsk.su},\, L.V.Il'ichov \\
{\it Institute for Automation and Electrometry SB RAS,} \\
{\it Novosibirsk 630090, Russia.}}
\date{} 
\begin{document}


\maketitle
\begin{abstract}
We report on a new phenomenon of molecular nuclear spin isomers conversion in the field of resonant laser radiation. It is based on the isomer-selective optical excitation and the difference of conversion rates for excited and unexcited molecules. We expect the considerable magnitude of the effect for all molecules which absorb the radiation of existing lasers.

\end{abstract}
PACS: 34.30+h\\

	In a recent work [1] a phenomenon of enrichment of nuclear spin isomers of molecules was predicted. The enrichment appears in the course of selective laser excitation of definite spin isomers. It was expected that the proposed enrichment method was the most applicable to the molecules where the nuclear spin conversion was induced by intramolecular spin-state mixing interaction. As it was shown in [1], the enrichment could be very high under an appropriate frequency of excited radiation. On the other hand, there were strict limitations placed on the laser frequency. In particular, the abilities of $CO_2$-laser and other sources appear to be insufficient for $CH_{3}F$ molecule, which is other than that a very convenient object for investigation of the enrichment phenomenon. The efficiently absorbed lines of $CO_2$-laser in $CH_{3}F$ do not satisfy the conditions from [1]. Thus the realization of the idea [1] is uneasy due to problems with compatible object and radiation source finding. 

	As it turns out the frequency constraints from [1] may be removed without any considerable loss of enrichment. There stays only one condition for the frequency to meet: The radiation should effectively be absorbed on a vibrational or electronic transition from the molecule ground state. This indulgence radically expands the set of relevant objects (those, demonstrating the light-induced conversion with available radiation sources).

	The physical nature of the proposed phenomenon is quite similar to that of light-induced drift (LID) [2]\footnote{In particular, the separation of nuclear spin isomers of molecules was firstly realized on the basis of LID [3]} and other light-induced gas-kinetic effects (see, e.g., [4]). Their common essence is in the following: if radiation excites particles selectively with respect to a physical parameter, than the difference of rates of relaxation of the parameter for excited and unexcited particles shifts the mean stationary value of the parameter from the equilibrium. 

	This is the definite type of nuclear spin isomers, which plays the role of the mentioned parameter. The concentration of the isomer gives the numerical mean value of the parameter. Radiation excites molecules selectively with respect to their nuclear spin modification. The conversion rates are generally different for excited and unexcited molecules in any conversion scenario. The difference should be the most dramatic when the conversion is caused by intramolecular magnetic interaction. This conversion mechanism, which was unambiguously proved for $CH_{3}F$ [5-12] and which seemingly works in other heavy and complicated molecules, has a resonance nature: the conversion rate increases drastically when the quantum state of mixing spin isomers are nearly degenerate. In the set of all populated rotational levels of a given spin isomer the resonance condition may distinguish very few or even a single one. Through this "most resonant" level the conversion takes place. The realization of identical resonance conditions for rotational levels of various vibrational (electron) states is unfeasible, which leads to the difference of conversion rates in these states. In what follows we will keep in mind the mechanism of conversion induced by intramolecular mixing.

	For simplicity we assume that there are only two nuclear spin modifications: ortho- and para-molecules (see fig.1).
\begin{figure}[hbt]
\centerline{\psfig{figure=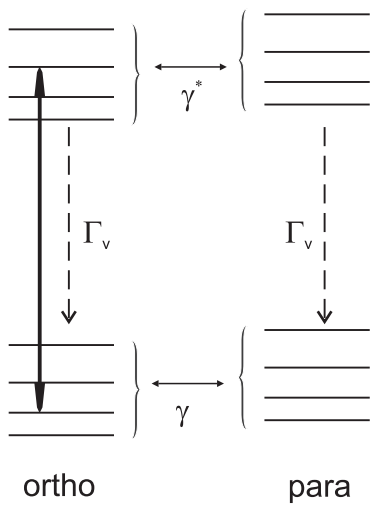}}
\caption{Illustration of the light-induced conversion phenomenon. Solid arrow indicates the transitions induced by radiation.}
\end{figure} 
Let the laser radiation interact with ortho-isomers only. The population distribution over rotational levels is assumed equilibrium in every vibrational state. The last assumption is justified when the vibrational relaxation is sufficiently slow in comparison with the rotational one, which can be provided by suitable buffer gas admixing. Under these condition we have   
\begin{eqnarray}
\frac{dN_o}{dt} = \gamma^{\ast}(N_{p}^{\ast} - N_o^{\ast}) + \gamma[(N_p - 
N_{p}^{\ast}) - (N_o - N_{o}^{\ast})] \nonumber \\
=(\gamma^{\ast} - \gamma)(N_{p}^{\ast} - N_{o}^{\ast}) + \gamma(N_p - N_o);
\end{eqnarray}
$$
\gamma^{\ast}  \equiv  \sum_{J_{o}^{\ast}}\gamma_{op}^{\ast}W(J_{o}^{\ast}),\;\; \gamma \equiv \sum_{J_o}\gamma_{op}W(J_o).
$$
Here $N_o$, $N_p$ are, respectively, the concentrations of ortho- and para-isomers; $\gamma$, $\gamma^{\ast}$ are the conversion rates. The asterisk indicates quantities for excited vibrational state. $\gamma_{op}W(J_o)$ gives the partial conversion rate through the rotational $J_o$-level of ortho-isomers. $W(J_o)$ is the Boltzmann factor.

	It follows from (1) that in the absence of radiation ($N_{p}^{\ast} = N_{o}^{\ast} = 0$) and under stationary condition, one has $N_p = N_o$. This is the thermodynamic equilibrium. The equilibrium will not be disturbed by radiation, if $\gamma^{\ast} = \gamma$. In the other case ($\gamma^{\ast} \neq \gamma$) a new stationary situation takes place:
\begin{equation}
N_p - N_o = (1 - \frac{\gamma^{\ast}}{\gamma})(N_{p}^{\ast} - N_{o}^{\ast}).
\label{2}
\end{equation}
Although only ortho-isomers are excited by the laser beam, there will appear excited para-isomers ($N_{p}^{\ast}\neq0$) due to resonant excitation transmission. The conversion is the slowest relaxation process in our system. Hence the level populations instantly follow the current concentrations of ortho- and para-isomers. In particular, for excited para-isomers the following balance equation is valid:  
\begin{equation}
\alpha N_{o}^{\ast}(N_p - N_{p}^{\ast}) - \alpha N_{p}^{\ast}(N_o -N_{o}^{\ast}) = \Gamma_{v}N_{p}^{\ast}.
\label{3}
\end{equation}
Here the first term in the lhs describes the creation of excited para-isomers due to excitation exchange with ortho-isomers; the second term corresponds to the reverse process. The coefficient
$\alpha$ gives the excitation exchange rate. The term in the rhs of (3) is conditioned by the decay of excited vibrational state. As it follows from (3)
\begin{equation}
N_{p}^{\ast} = \frac{\alpha N_p}{\Gamma_v + \alpha N_o}N_{o}^{\ast}
\label{4}
\end{equation}
By the proper choice of buffer gas and its pressure, one can guarantee
\begin{equation}
\Gamma_v \gg \alpha (N_o + N_p),
\label{5}
\end{equation}
Then $N_{p}^{\ast}\ll N_{o}^{\ast}$, and the quantity $N_{p}^{\ast}$ in (2) may be neglected. In this case the deviation of ortho- and para-concentrations from the equilibrium is maximal and most simply related to the fraction of excited ortho-isomers. It is known that under specific conditions (sufficiently high radiation intensity, proper relations between relaxation rates) this fraction can rich 1/2. Hence it follows from (2) that  
\begin{equation}
\frac{N_p}{N_o} =\frac{1}{2}(1 + \frac{\gamma^{\ast}}{\gamma}).
\label{6}
\end{equation}
Two limiting cases of this expression are clearly interpretable. If $\gamma^{\ast}\ll\gamma$, then $N_o = 2N_p$ because the conversion hardly goes through the excited vibrational state, whereas the conversion through the ground state leads to equalization of ortho- and para-concentrations in this state. Remember that the concentration of ortho-isomers in the ground state is the half of their total concentration. In the opposite limiting case ($\gamma^{\ast}\gg\gamma$) we have $N_o \ll N_p$. That is, almost all molecules are converted into para-modification due to conversion through the excited state. Note that the reverse process is suppressed because para-isomers are for the most part in the ground state.

	Let us discuss the conditions which provide the rotational distributions in vibrational states to be close to the Boltzmann ones. If the laser radiation is absorbed on a rovibrational transition, the relative excess of the population of the upper level $J_L$ over the  amplitude of the Boltzmann "background" is determine by the parameter    
(see, e.g., [4,13])
\begin{equation}
a = \frac{\Gamma_v}{\Gamma W(J_L)},
\label{7}
\end{equation}
where $\Gamma$ is the impact line halfwidth, which is mainly conditioned by the rotational relaxation (the broadening is assumed homogeneous); $W(J_L)$ is the Boltzmann factor for the level $J_L$. The parameter $a$ can be small even for the pure gas of converting molecules. But it can be made even smaller by admixing a buffer gas with the cross-section of the vibrational state deexcitation being small in comparison with the broadening cross-section. When $a\ll 1$ the distribution of population over rotational levels may safely be approximated by the Boltzmann distribution.

	 The total population of the excited vibrational state of ortho-molecules is given in this case by the relation [4,13]
\begin{equation}
N_{o}^{\ast}  =  N_o\frac{\kappa/2}{1 + \kappa + 
\Omega^2/\Gamma^2};\;\; 
\kappa = \frac{4|G|^2}{\Gamma_{v}\Gamma}W(J_L);\;\; G = \frac{Ed}{2\hbar}.
\label{8}
\end{equation}
Here $\Omega$ is the frequency detuning; $E$ is the amplitude of the radiation field; $d$ is the dipole moment of the resonant transition; 
$\kappa$ is the so-called saturation parameter. It is evident from (8) that for high saturation parameter one has $N_{o}^{\ast} = N_{o}/2$, which was used in (6).

	Let us estimate the possible value of the effect for
${}^{13}CH_3 F$-molecules, which are in good resonance ($\Omega\simeq$ 26 MGz 
with the Doppler width$\simeq$ 40 MGz) with 
$CO_2$ laser (the line $P(32)$ of the band 9,6 $\mu$) on the transition $R(4,3)$. In this case only ortho-isomers absorb the radiation. We assume the absorptive cell has the length $\simeq 30$ cm. Taking into account that the absorption coefficient 
 $\sim 0,3$ cm$^{-1}$/Torr [14], the probe testing laser beam is efficiently absorbed on this length, when the pressure of ${}^{13}CH_3 F$ is about 0.1 Torr. For this value of ${}^{13}CH_3 F$ pressure the forthcoming estimations are done.
The condition (5) is not met without special cares. For the increase of $\Gamma_v$ we assume the addition of a buffer gas of molecules with proper vibrational quantum, so that the vibrational excitation can efficiently be transmitted from ${}^{13}CH_3 F$. To make this process irreversible, it is worth to take the buffer gas with a bit smaller vibrational quantum. Assuming the rate of excitation transfer from ${}^{13}CH_3 F$ to the buffer molecules to be comparable with the excitation exchange between two ${}^{13}CH_3 F$-molecules and making this buffer gas pressure 0.5 Torr, we get $\Gamma_v  = 5 \alpha (N_o + N_p)$. Thus, the condition (5) is fulfilled.

	Let us consider now the parameter (7). According to the data [15] on the rate of excitation exchange between 
${}^{13}CH_3 F$ and ${}^{12}CH_3 F$ 
($5\cdot 10^5$ s$^{-1}$/Torr), we get  
 $\Gamma_v \cong 40$ KGz. For the total gas pressure and for the broadening caused by the resonant buffer gas ($\simeq 20$ MGz/Torr [16]), we have $\Gamma \simeq 12$ MGz. Taking into account that $W(J_L)\sim 10^{-2}$, the parameter $a\simeq 1/3$. This value is not sufficiently small. If, however, we add 10 Torr of $He$ as a second (nonresonant) buffer gas, then its broadening effect (3 MGz/Torr [16]) gives $\Gamma\cong 40$ MGz. On the contrary, the value of $\Gamma_v$ stays almost untouched because the deexcitation effect of the helium ($\sim 10^2$ Gz/Torr [17]) is negligible. Under these conditions we have 
$a\simeq 1/10\ll 1$, i.e., (7) is fulfilled. Note that the obtained value of $\Gamma$ is approximately equal to the Doppler halfwidth. So, the estimations, which are based on the homogeneous broadening assumption, are correct.

	Finally, let us estimate the attainable saturation parameter from (8). For the radiation power density
 $\sim 10^3$ W/cm$^2$ (the laser beam of 1 mm in diameter and 10 W of power) we have $\kappa\simeq 50$ in accordance with [14]. This guarantee the saturation of the vibrational transition.

	Hence the above suggested experimental conditions are close to the optimum for the manifestation of the light-induced conversion effect. The considered simple expression (6) is quite suitable for estimations.

	It is worth to note that the fulfillment of (5) is not indispensable at least in the case
$\gamma^{\ast}\gg\gamma$, when one should expect the brightest manifestation of the phenomenon. More detailed estimations show that even for $\Gamma_v \simeq \alpha (N_o + N_p)$ the almost complete conversion from ortho- to para-isomers takes place, provided  $N_o^*$ is not too small in comparison with $N_o$.

	There is one further circumstance: The external electric field can radically effect the conversion rate and as a consequence the stationary state. This is due to the Stark shift of levels which the conversion goes through (both in the ground and excited states). The levels can so be put into (or out of) resonance. The influence of the external electric field on the conversion of nuclear spin isomers of $CH_3 F$ was  rigorously proved and investigated in [10,11].
There  follows a conclusion from these works that the electric field method of identification of the converting levels is promising.

	Summarizing, we shown that in the system of nuclear spin isomers of molecules where only one isomer absorbs the resonant radiation the simple experimental conditions can provide the shift of stationary concentrations of isomers from equilibrium concentrations. There can be almost total exhaustion of one isomer.

	The work was supported in part by the program RFBR (grants 98-02-17924 and 98-03-33124a).

 

\begin{thebibliography}{07}
\bibitem[1]{1}
L.V.Il'ichov, L.J.F.Hermans, A.M.Shalagin, P.L.Chapovsky, {\it Chem.Phys.Lett.} {\bf 297}, 439 (1998).
\bibitem[2]{2}
F.Kh.Gel'mukhanov, A.M.Shalagin, {\it JETP Lett.}\, {\bf 29}, 711 (1979).
\bibitem[3]{3}
L.N.Krasnoperov, V.N.Panfilov, V.P.Strunin, and P.L.Chapovsky, {\it Pis'ma Zh.Eksp.Teor.Fiz.}\, {\bf 39}, 143 (1984).
\bibitem[4]{4}
S.G.Rautian, A.M.Shalagin, {\it Kinetic Problems of Non-linear Spectroscopy}, North-Holland, 1991.
\bibitem[5]{5}
P.L.Chapovsky, {\it JETP}\, {\bf 97}, 895 (1990).
\bibitem[6]{6}
P.L.Chapovsky, {\it Phys.Rev. A} {\bf 43}, 3624 (1991).
\bibitem[7]{7}
B.Nagels, M.Schuurman, P.L.Chapovsky, L.J.F.Hermans, {\it J.Chem.Phys.} {\bf 103}, 5161 (1995).
\bibitem[8]{8}
B.Nagels, M.Schuurman, P.L.Chapovsky, L.J.F.Hermans, {\it Phys.Rev. A} {\bf 54}, 2050 (1996).
\bibitem[9]{9}
P.L.Chapovsky, {\it Physica A} {\bf 233}, 441 (1996).
\bibitem[10]{10}
B.Nagels, M.Schuurman, L.J.F.Hermans, P.L.Chapovsky, {\it Chem.Phys.Lett.} {\bf 242}, 48 (1995).
\bibitem[11]{11}
B.Nagels, N.Calas, D.A.Roozemond, L.J.F.Hermans, P.L.Chapovsky, {\it Phys.Rev.Lett.} {\bf 77}, 4732 (1996).
\bibitem[12]{12}
E.Ilisca, K.Bahloul, {\it Phys.Rev. A} {\bf 57}, 4296 (1998).
\bibitem[13]{13}
V.R.Mironenko, A.M.Shalagin, {\it Bull.Acad.Sci.USSR, Phys.Ser.}
\, {\bf 45}, 87 (1981).
\bibitem[14]{14}
V.N.Panfilov, V.P.Strunin, and P.L.Chapovsky, {\it Zh.Eksp.Teor.Fiz.}\, {\bf 85}, 881 (1983).
\bibitem[15]{15}
E.N.Chesnokov, V.N.Panfilov, {\it Zh.Eksp.Teor.Fiz.}\, {\bf 73}, 2122 (1977).
\bibitem[16]{16}
W.K.Bihel, P.J.Kelly, Ch.K.Rhodes, {\it Phys.Rev. A} {\bf 13}, 1817 (1976).
\bibitem[17]{17}
E.Weitz, G.W.Flynn, {\it J.Chem.Phys.} {\bf 58}, 2678 (1973).
\end{thebibliography}
\end{document}